\documentclass[english]{article} 
\usepackage{babel}
\usepackage{natbib}
\usepackage[T1]{fontenc} 
\usepackage[latin9]{inputenc} 
\usepackage[letterpaper]{geometry}  
\usepackage[]{algorithm2e}
\usepackage{framed}
\usepackage{graphicx} 
\usepackage{esint} 
\usepackage{setspace}
\usepackage{authblk}

\begin{document}
	
	{
	\setstretch{1.0}
	\title{Inferring subsurface heterogeneity from push-drift tracer tests\footnote{Los Alamos National Laboratory technical report: \textbf{LA-UR-17-21910}}}
		
	
	\author[1]{Scott K. Hansen}
	\author[1]{Velimir V. Vesselinov}
	\author[2]{Paul W. Reimus}
	\author[1]{Zhiming Lu}
	
	\affil[1]{Computational Earth Science Group, Los Alamos National Laboratory, Los Alamos, NM, USA}
	\affil[2]{Earth System Observations Group, Los Alamos National Laboratory, Los Alamos, NM, USA}
	
	\date{March 28, 2017}
	\renewcommand\Affilfont{\itshape\small}
	
	\maketitle
	
	\begin{abstract} 
		We consider the late-time tailing in a tracer test performed with a push-drift methodology (i.e., quasi-radial injection followed by drift under natural gradient). Numerical simulations of such tests are performed on 1000 multi-Gaussian 2D log-hydraulic conductivity field realizations of varying heterogeneity, each under eight distinct mean flow directions. The ensemble pdfs of solute return times are found to exhibit power law tails for each considered variance of the log-hydraulic conductivity field, $\sigma^2_{\ln K}$. The tail exponent is found to relate straightforwardly to $\sigma^2_{\ln K}$ and, within the parameter space we explored, to be independent of push-phase pumping rate and pumping duration. We conjecture that individual push-drift tracer tests in wells with screened intervals much greater than the vertical correlation length of the aquifer will exhibit quasi-ergodicity and that their tail exponent may be used to infer $\sigma^2_{\ln K}$. We calibrate a predictive relationship of this sort from our Monte Carlo study, and apply it to data from a push-drift test performed at a site of approximately known heterogeneity---closely matching the existing best estimate of heterogeneity.
	\end{abstract}
	}
	\section{Introduction}
		Single-well injection withdrawal (SWIW) or \textit{push-pull} tracer tests are commonly performed, and interpretation methodologies have been proposed for inference of many subsurface parameters from their breakthrough curves. This body of interpretive theory generally assumes radially-symmetric flow, with particles tracing the same paths on their outbound and inbound journeys and implying last-in-first-out (LIFO) behavior. (The literature review in \cite{Hansen2016} contains an extensive discussion of these assumptions.) Background groundwater flow (drift) and heterogeneity cause non-radial, hysteretic flow patterns and violate the LIFO assumption, complicating test interpretation by generating heavy tails that may spuriously be attributed to other causes \citep{Lessoff1997}. \cite{Johnsen2009a} presented analysis quantifying the effect of solute path hysteresis caused by background drift on push-pull breakthrough curves obtained from homogeneous velocity fields and \cite{Hansen2016} studied this phenomenon numerically in heterogenous velocity fields.
		
		Comparatively little has been published which exploits the asymmetry between injection and extraction phases caused by background drift in order to extract information about the subsurface. Exceptions include the works of \cite{Leap1988} and \cite{Hall1991}, who studied use of drift-pumpback (``push'' phase under natural gradient and pull phase under forced gradient) to measure groundwater velocity. Also, \cite{Novakowski1998} presented an analytical-numerical approach to inference of matrix diffusions from push-drift tests (push phase under forced gradient, ``pull'' phase under natural gradient) in fractured porous media. All three of these approaches assumed idealized (radial or linear) flow regimes.  
		
		In this note, we further consider the information that can be obtained from an asymmetric, push-drift, test, but exploit the basic numerical framework developed by \cite{Hansen2016} to study the behavior of such tests in heterogeneous media. In particular, our goal is to \textit{assess} the degree of heterogeneity from push-drift test breakthrough curves.
		
\begin{figure}
			\centering
			\includegraphics[scale=0.7]{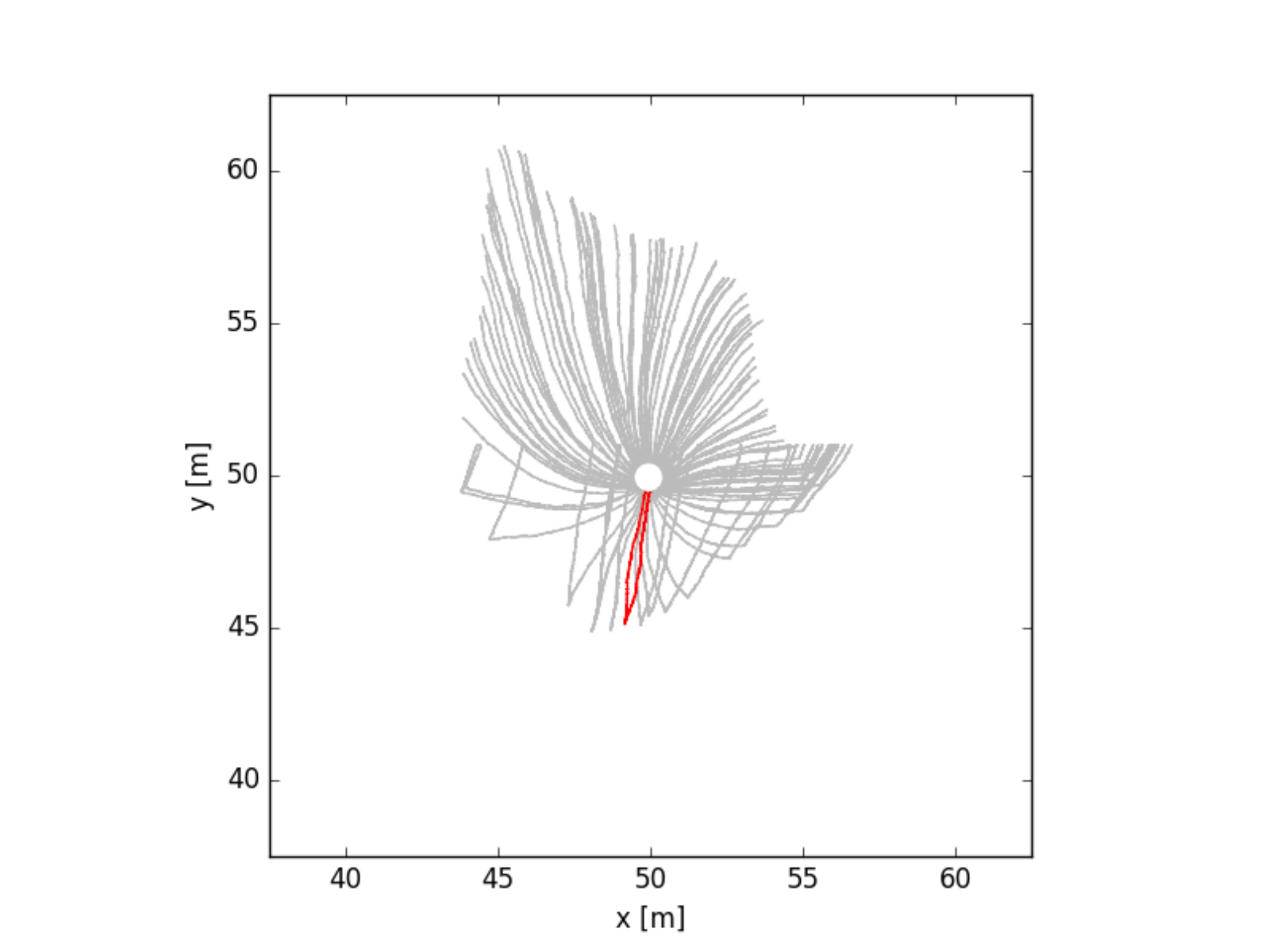}
			\caption{Particles tracks from a push-drift simulation with 100 particles tracked in a realization with $\sigma^2_{\ln K}=1.5$, and mean background drift in the $y$-direction. Particle paths are traced during the push phase, and if at any time during the drift phase are noted to be significantly downgradient of the injection ring, they are frozen. Paths that never return to the injection ring are gray, the path that eventually returns is highlighted in red.}
			\label{fig: paths}
		\end{figure}

		Our motivation for this attempt is as follows: In a push-drift test in a 2D hydraulic conductivity field ($K$-field), only a small packet of solute leaving the well at time zero will return to the well, with all other solute packets missing the well during the drift phase, as seen in Figure \ref{fig: paths}. 
		In a totally homogeneous aquifer, streamlines in both radial flow and in background drift are straight, so (excluding local-scale dispersion) there is essentially no path line hysteresis experienced by particles recaptured at the well. 
		In heterogeneous media, the packet of solute that ultimately returns to the well will generally travel along disjoint outbound and inbound streamlines, each with different velocities.
		Since the distribution of streamline velocities increases with heterogeneity, it is reasonable to posit that the distribution of return times in an ensemble of 2D $K$-fields with the same heterogeneity statistics will similarly broaden with increasing subsurface heterogeneity, and will thus contain information about heterogeneity.
		Furthermore, under the common horizontal flow assumption for layered aquifers \citep[e.g.,][]{Pickens1981,Guven1985,Klotzsch2016a} the observed breakthrough curve is a flux-weighted average of the breakthrough curves for multiple layers intersected by the screen, each of which may be conceived of as an independent 2D $K$-field realization. Assuming no correlation between hydraulic conductivity at the well bore and path-line hysteresis, as well as no pore-scale dispersion, we conclude that the push-drift breakthrough curve will be approximated by the return time pdf for an ensemble of 2D $K$-field realizations.
		
		Consequently, we aim to relate properties of the ensemble 2D return time pdf to $\sigma^2_{\ln K}$. Because of the difficulties inherent in approaching such stochastic problems analytically, we perform a large-scale Monte Carlo computational study of push-drift tests to gain insight into this relationship. In Section 2, we describe the Monte Carlo study. In Section 3, we fit establish an empirical relationship between heterogeneity and late-time push-drift breakthrough behavior. In Section 4, we corroborate our relation against a real data set. In Section 5, we recap what we have learned.
		
	\section{Methodology of numerical study}
		We performed a computational study, using 1000 realizations of 50 m by 50 m, multi-Gaussian, isotropic 2D log hydraulic conductivity fields, with constant conductivities assigned to each cell on a 100 by 100 grid. All realizations assumed an exponential semivariogram with a correlation length of 4 m, and a geometric mean hydraulic conductivity of $10^{-4}$ m/s. The realizations were divided into batches of 250, each batch featuring a different value of $\sigma^2_{\ln K}$, respectively: 0.5, 1.0, 1.5, and 2.0.
		
		For each conductivity field, with index $i$ we ran three steady-state flow simulations in PFLOTRAN \citep{lichtner2015pflotran}. For the first (quasi-radial) simulation, we imposed a constant mass injection rate per unit depth of 0.1 kg/m/s at the center and zero head at all points on the outer boundary, and computed a velocity field, $\mathbf{R}_i$, which is represented as 100 by 100 by 2 tensor of  $x$- and $y$-direction cell center velocities. For a second simulation, we imposed no flow boundary conditions on the north and south faces (i.e. at $y=25$ and $y = 75$), and constant head values, higher at the west edge ($x = 25$), and lower at the east ($x = 75$). The resulting velocity field, $\mathbf{X}_i$, with mean flow in the $x$-direction, was computed, and then normalized to give a mean flow velocity of 0.01 m/h. For a third simulation, we imposed no flow boundary conditions on the west and east faces (i.e. at $x=25$ and $x = 75$), and constant head values, higher at the south edge ($y = 25$), and lower at the north ($y = 75$). The resulting velocity field, $\mathbf{Y}_i$, with mean flow in the $y$-direction, was computed, and then normalized to give a mean flow velocity of 0.01 m/h. Since tracer only interrogates the area immediately surrounding the well, the no-flow boundary conditions imposed at the north and south edges of the domain for the quasi-linear simulations were not considered to be relevant. In all three cases, velocity fields were steady-state. These velocity fields were used to simulate push-pull tests under a variety of conditions.
		
		\begin{figure}
			\includegraphics{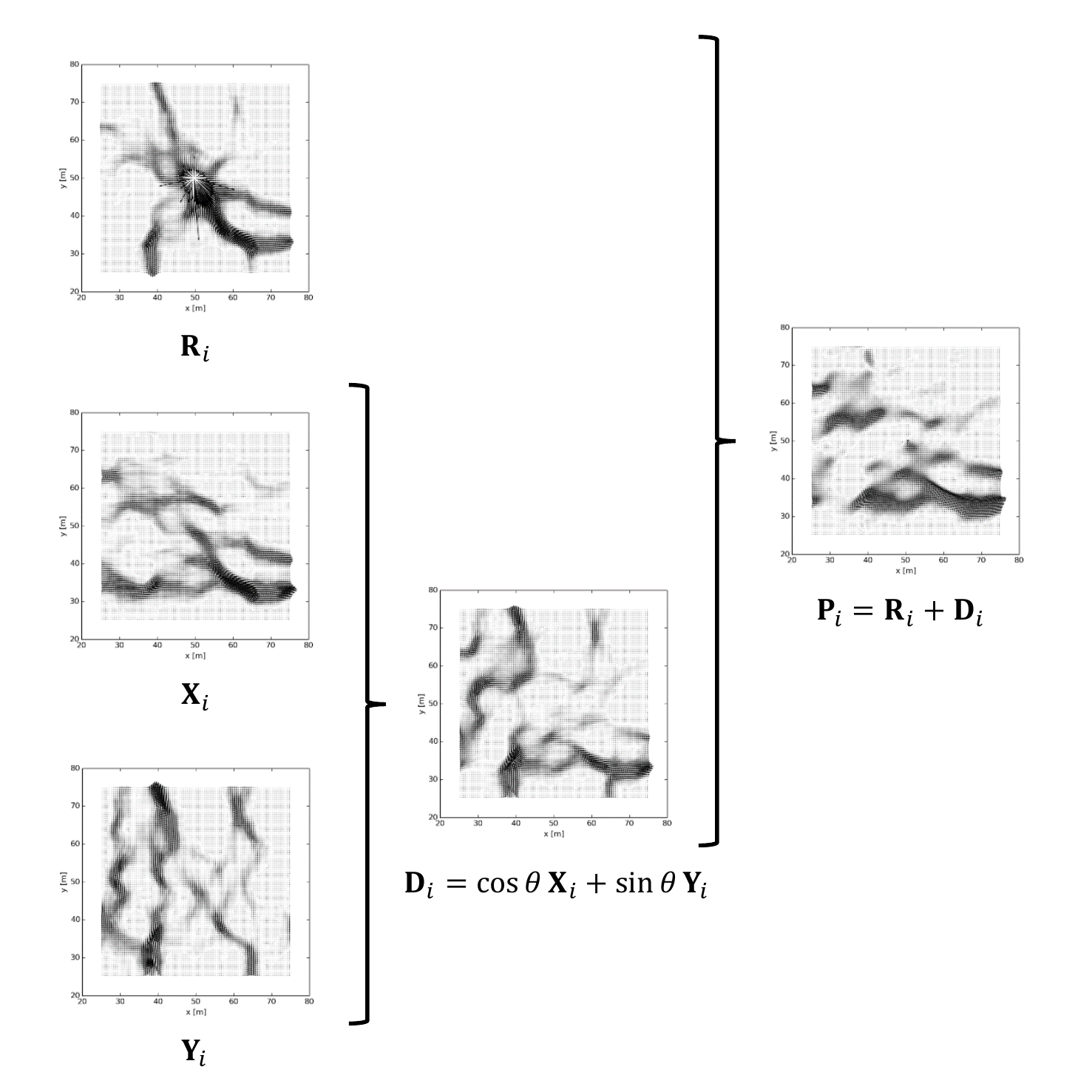}
			\caption{Visual representation of the superposition of radial, $\mathbf{R}_i$, $x$-direction, $\mathbf{X}_i$, and $y$-direction, $\mathbf{Y}_i$, velocity fields to generate the drift phase vector field, $\mathbf{D}_i$, and push phase vector field, $\mathbf{P}_i$, for a single realization, $i$. Quiver plots for each field are shown, with braces indicating a superposition operation. In this example, $\sigma^2_{\ln K}=1.5$ and $\theta=\pi/4$. Note that arrow length scale is not the same between different quiver plots.}
			\label{fig: composition}
		\end{figure}
		
		For each $K$-field realization, eight push-phase and corresponding drift-phase velocity fields were generated. These were determined in identical fashion, except for the direction of mean flow (all of the cardinal and inter-cardinal directions were examined for each realization). Using the principle of superposition, a linear combination of the velocity fields from the three PFLOTRAN flow simulations described above was computed to determine the velocity field during both the push and the drift phases of the test. Where the angle of mean flow to the x-axis is $\theta$, background drift was simulated for each realization by scaling all the cell center velocity vectors from the mean $x$-direction simulation by $\cos(\theta)$ and all the cell center velocity vectors from the mean $y$-direction simulation by $\sin(\theta)$. The drift phase velocity tensor is computed as $\mathbf{D}_i = \cos(\theta) \mathbf{X}_i + \sin(\theta) \mathbf{Y}_i$ and the push phase velocity tensor as $\mathbf{P}_i = \mathbf{R}_i + \mathbf{D}_i$. This superposition process is illustrated in Figure \ref{fig: composition} for one particular realization. The corresponding scaled vectors for each cell were added to generate the effective background drift velocity in each cell. In all cases the average background drift velocity had magnitude 0.01 m/h; this velocity was chosen arbitrarily (its magnitude scales the distribution of particle return times, but does not alter the power law exponent of this distribution). During the push phase, the corresponding cell center velocity vectors from the quasi radial simulation were also added.
		 
		For each of the eight sets of push-drift velocity fields, for each $K$-field realization, three distinct push-drift particle tracking simulations were performed, of types A, B, and C. Each employed constant, small time steps of duration 0.05 h. For each particle, at each time step, the velocity field was interpolated based on the particle's starting location. For the entire time step, the particle travels along its local streamline, and then undergoes a small random Fickian dispersive motion determined by $\alpha_l = 0.01$, $\alpha_t = \alpha_l/10$, and the streamline velocity. The characteristic pore scale dispersivity, $\alpha_l$, was chosen based on the reported ranges in \citet{Schulze-Makuch2005} for a sandy aquifer. All particle tracking simulations commenced by introducing particles in a ring around the injection location at the center of the domain. In type A simulations, particles were tracked during a ``slow'' push phase of 80 h, during which radial and background flow were operative, generating a characteristic interrogation radius of 3.0 m. In type B simulations, the push phase was taken to be instantaneous (i.e., the same volumetric injection of water was simulated as in type A simulations, but no background drift was operative). type C simulations featured the same ``slow'' injection rate as in type A, but featured a push phase that lasted 160 h, so twice as much chase water was injected, leading to a characteristic interrogation radius of 4.3 m. In all simulations, the subsequent drift phase had identical physics. This proceeded for 4000 h for type A and B simulations, and 5000 h for type C simulations, or until all particles had passed or entered the injection ring. No processes other than local-scale dispersion and advection affected the particles. 
		
		For each push-drift simulation (6000 simulations were performed for each value of $\sigma^2_{\ln K}$: eight flow directions times 250 distinct $K$-field realizations times three push-phase implementations), if a particle re-entered the ring during the drift phase of the simulation, the time at which this occurred was recorded and the particle was removed from the system. At the end of each simulation the average time of arrival was recorded. An example simulation is shown in Figure \ref{fig: paths}.

	\section{Relationship between $\sigma^2_{\ln K}$ and power law exponent}
\begin{figure}
			\centering
			\includegraphics[scale=0.6]{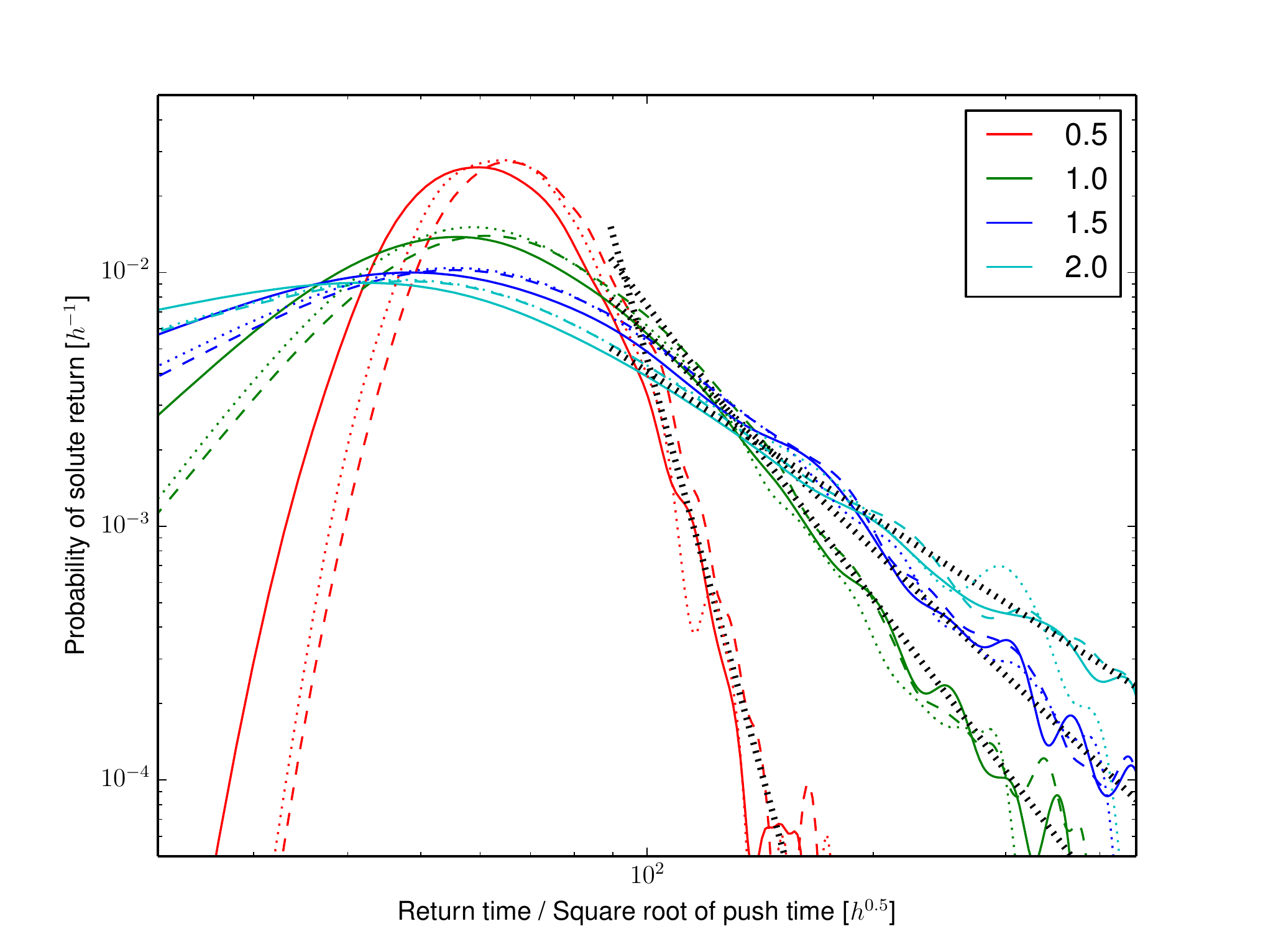}
			\caption{Log-log plot of ensemble return time pdfs for each of four values of $\sigma^2_{\ln K}$: 0.5, 1.0, 1.5, and 2.0. Solid, dashed, and dotted curves derive, respectively, from simulations of types A, B, and C. The power law tails fit by linear regression are superimposed as thick, black dotted lines.}
			\label{fig: btcs}
		\end{figure}

		For each of the $\sigma^2_{\ln K}$, for each push-phase regime (type A, B, or C), Gaussian kernel density estimation was applied to generate pdfs from the return times derived from the 2000 simulations Since each pdf was determined to have a power law tail, the tails were plotted in log-log space and linear regression was applied to determine the tails' slopes. The empirical pdfs (normalized by square root of the push phase time for easy comparison) and their superimposed regression lines are presented in Figure \ref{fig: btcs}. We note that the tail slope is not seen to be affected by injection rate or injection volume, within the limits we explored. Naturally there will be some characteristic interrogation radius that is too small to properly sample the domain, but we conclude that it is below the radii considered here, which are of the same order as the correlation length of the $K$ fields.
		
\begin{figure}
 			\centering
 			\includegraphics[trim={2cm, 2cm, 2cm, 2cm}, clip ,scale=0.5]{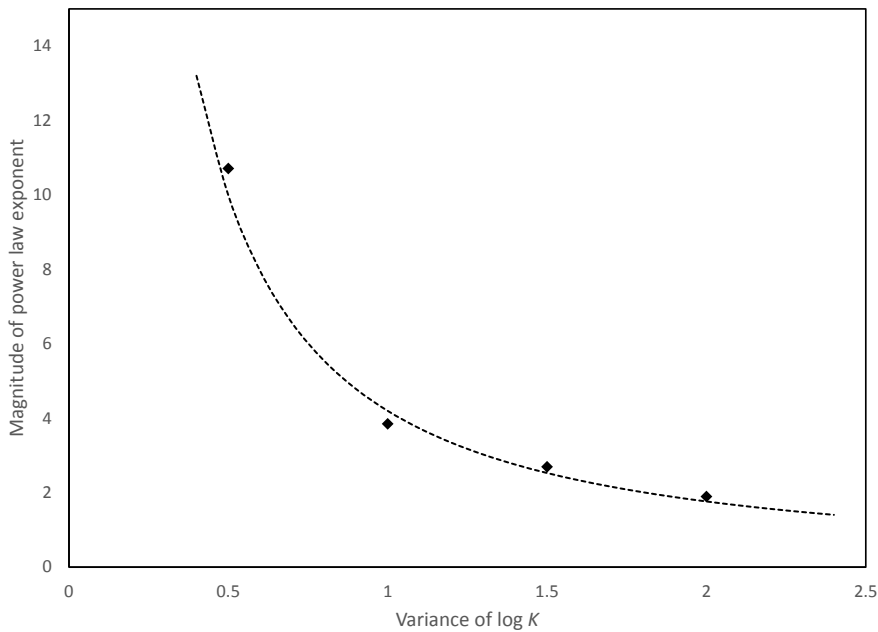}
 			\caption{Dashed line: plot of $\sigma^2_{\ln K}$ against $|\alpha|$, according to (\ref{eq: relation}). Diamonds: actual slopes of late-time tails in log-log space shown in Figure \ref{fig: btcs}, as determined by the Monte Carlo study.}
 			\label{fig: relation}
 		\end{figure}

		 For each of $\sigma^2_{\ln K}=$ 0.5, 1.0, 1.5, and 2.0, late-time slopes of $-10.7$, $-3.8$, $-2.7$, and $-1.9$, respectively, were determined. Since the slopes are numerically equal to the corresponding power law exponent, these data lead to the simple empirical approximation
		 \begin{eqnarray}
			 c(t) &\propto& t^{-\alpha},\label{eq: exponent}\\
			 \alpha &\approx& 4.2\ \left(\sigma^{2}_{\ln K}\right)^{-1.25},
			 \label{eq: relation}
		 \end{eqnarray}
		 for large $t$, where $t$ represents the time since test commencement, and $c(t)$ represents the concentration observed at the well. The regression slopes, along with $|\alpha|$ determined by (\ref{eq: relation}) are shown in Figure \ref{fig: relation}.
		 
	\section{Corroboration of concept using real data}
\begin{figure}
			\centering
			\includegraphics[trim={2cm, 2cm, 2cm, 2cm}, clip ,scale=0.5]{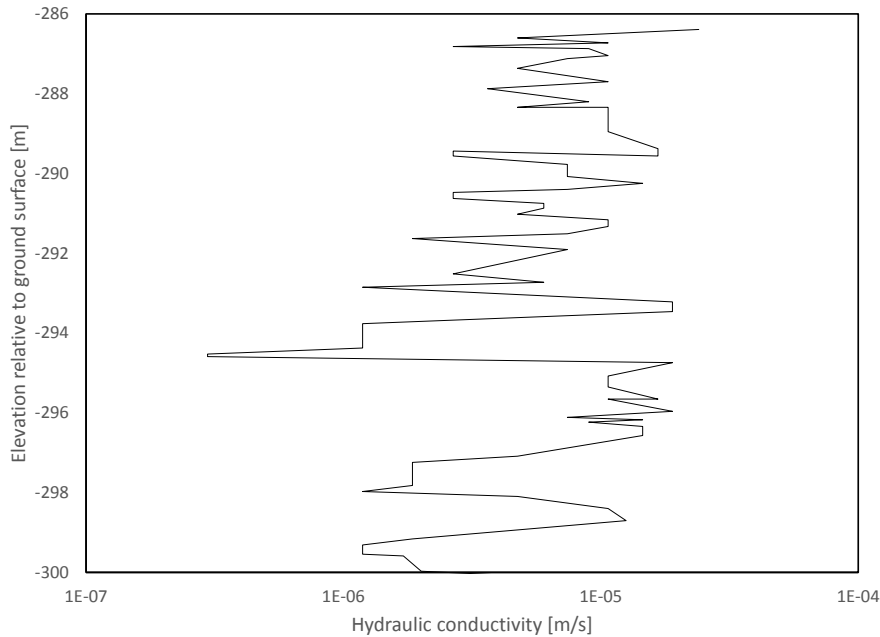}
			\caption{Plot of hydraulic conductivity profile in the sedimentary unit in which the push-drift tracer test was performed, estimated from direct examination of core obtained in the vicinity of the well screen at which the push-drift test was performed and application of Kozeny-Carman correlation.}
			\label{fig: CH-2}
		\end{figure}

		To demonstrate the relation we have developed, we test it against a push-drift data set obtained from a well completed in the regional aquifer beneath Mortandad Canyon on the grounds of the Los Alamos National Laboratory. The portion of the aquifer in which the push-drift tracer test was performed has been relatively well characterized, and a complete sedimentary record was obtained from a nearby core hole for the sedimentary unit in which the well has been screened. Using the Kozeny-Carman correlation \citep{Carman1997}, it was possible to estimate the hydraulic conductivity profile at the core hole. This is shown in Figure \ref{fig: CH-2}. Based on this data, we conclude that $\sigma^2_{\ln K}\approx1$. 
		
\begin{figure}
			\centering
			\includegraphics[trim={2cm, 2cm, 2cm, 2cm}, clip ,scale=0.5]{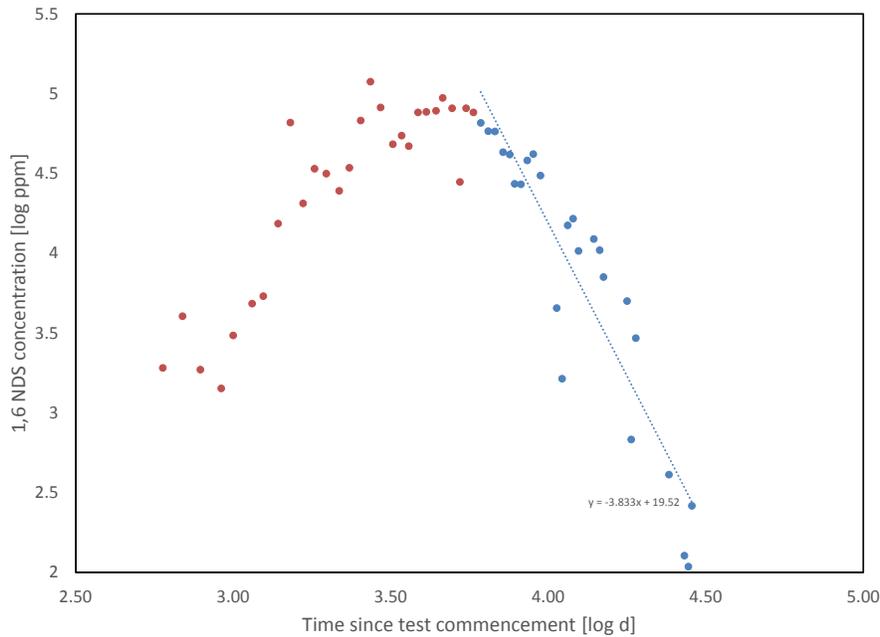}
			\caption{Log-log plot (base $e$) of experimental push-drift 1,6-NDS concentrations, along with linear trend line fit to breakthrough curve tail. Points used for tail regression are shown in blue.}
			\label{fig: R-28}
		\end{figure}

		The push-drift test was performed under confined conditions, in a well with a 11.43 cm inner diameter, 12.7 cm outer diameter casing, installed in a 31.1 cm diameter bore hole with a 7.25 m long screened section which was surrounded by a 13.41 m filter pack. The mean depth of the screen is 288.4 m and the mean depth of the filter pack is 288.9 m below ground surface. The water table varies seasonally and has dropped over time but is currently about 273 m below ground surface. In total, 56.78 $\mathrm{m^3}$ of water with an approximately 1.3 mM concentration of 1,6-NDS tracer was injected over 4 h, followed by an equal chase volume of water without 1,6-NDS, injected over the subsequent 4 h. No further injection or extraction at the well was performed for the subsequent 90 d, save for periodic chemical sampling at the well, and the tracer was allowed to drift back into the well under background drift conditions. The resulting breakthrough data is shown in Figure \ref{fig: R-28}. The slope of the late-time breakthrough tail was determined in log-log space to be -3.8. Employing (\ref{eq: exponent}-\ref{eq: relation}), we arrive at a prediction of $\sigma^2_{\ln K} = 1.01$, very close to value obtained by direct core examination.
	
	\section{Conclusion}
		Using a large-scale Monte Carlo study, we explored the ensemble push-drift return time pdf for an ensemble of 2D, multi-Gaussian $K$-field realizations with a range of heterogeneities spanning from $\sigma^2_{\ln K}=0.5$ to $\sigma^2_{\ln K}=2$. We noted a power law tail on the pdfs for all examined heterogeneities, noted that its insensitivity to push-phase methodology, and observed that it was possible to relate the tail exponent to the heterogeneity, which we encapsulated in the simple relation (\ref{eq: relation}). 
	
		Many aquifers are known to be anisotropic, with typically only short-range correlation of hydraulic conductivities in their vertical direction compared to the horizontal directions. A sufficiently deep well will penetrate multiple vertical correlation lengths and, on the horizontal flow assumption, a push-drift test in such a well will generate a breakthrough curve that may be conceived of as the flux-weighted superposition of a large number of independent, 2D horizontal push-drift tests. We thus conjectured that such tests performed in the field will exhibit breakthrough curves proportional to the 2D ensemble pdfs determined here. We demonstrated this approach for interpretation of a push-drift tracer test breakthrough curve in an aquifer for which heterogeneity data exist, and found that it closely matches estimates from existing data.
		
	\section*{Acknowledgment}
		The authors acknowledge the support of the LANL environmental programs. Experimental data used to generate the figures, simulated hydraulic conductivity fields, and source code used for the simulations are archived by the lead author.

\end{document}